\documentstyle[12pt]{article}
\topmargin - 1.5cm
\leftmargin - 1cm

\begin{document}

\begin{center}

{\large \bf  Toward the spectral zone control }  \\[1cm] 
{\sc B. N. Zakhariev and V. M. Chabanov \\[3mm]}
Bogoliubov Laboratory of Theoretical Physics, Joint Institute for Nuclear
Research, \\ 141980 Dubna, Russia \\ e-mail:  zakharev@thsun1.jinr.ru;
chabanov@thsun1.jinr.ru

\end{center}

\begin{abstract} The desired shifts of the 
boundaries of spectral allowed  zones of periodical systems are
demonstrated. In particular, the phenomenon of merging neighbor allowed 
zones is exhibited and its simple explanation is given. It is also shown 
how to change the additional fundamental spectral parameter, the degree of 
exponential solution growth, at arbitrary given energy points inside 
the forbidden zones. This allows one to control tunneling through fragments 
of periodic structures at energies belonging to spectral gap. All the 
results are based on the finite interval {\bf inverse} eigenvalue problem 
which provides us with {\it complete sets} of exactly solvable models. This 
is a radical extension (continuous !  multiplicity) in comparison to the 
famous finite-gap  models. 

\vspace{0.4cm}


\vspace{0.4cm}

\end{abstract}

The periodical structures represent an area of intensive research and 
diverse practical applications, e.g., in microelectronics. So it is 
important to extend as far as possible the class 
of spectral zone control algorithms. We suggest, in particular, potential 
transformations leading to given shifts of chosen zone boundaries.
  
We apply our experience (quantum intuition) in  the {\it finite interval} 
inverse problem (IP)  algorithms [1-4], which allow the construction 
of the  potentials corresponding to a given set of spectral parameters, 
e.g., bound state energy levels and spectral weights $\{E_{\lambda }, 
c_{\lambda }\}$, {\it to periodic case}.  
 In the IP formalism [5-8] these parameters are input 
parameters uniquely determining the potential and so can be considered as 
spectral "control levers".  Corresponding {\bf exact models form the 
complete sets}.  The next step is the periodical continuation over the 
whole axis the potential derived with the inversion 
procedure on finite interval.  So we use the missed possibility to combine 
{\it inverse} problem approach on a finite interval with {\it direct} one 
of constructing solutions for the periodically continued potential 
(infinitely enriched Kronig-Penney-like procedure).  This gives us 
additional flexibility of the formalism in comparison with the {\it pure 
 inverse} finite gap theory of spectral band structures.

Let us use, for example, the formulae for energy level shift in 
infinite rectangular potential well of finite width. Let 
$\psi_{0}(x,E_{n})$ be corresponding eigenfunction at the energy $E_{n}$. 
We assume ${\bar \psi}_{0}(x,E_{n}+t)$ to be a non-physical auxiliary 
solution in the initial potential at the shifted energy $E_{n}+t$ with the 
symmetry being opposite to one of the $\psi_{0}(x,E_{n})$.  With these 
solutions, we can construct the Wronskian \begin{eqnarray} \theta(x) 
\enskip = \enskip \psi_{0}'(x, E_{n}) {\bar \psi}_{0}(x,E_{n}+t) \nonumber 
\\ - \psi_{0}(x, E_{n}) {\bar \psi}_{0}'(x,E_{n}+t).  \end{eqnarray} The 
final expressions for the transformed potential and solutions are
\begin{eqnarray} V_{2}(x) \enskip = \enskip \\ V_{0}(x) - 2 \, t \, 
\frac{d}{dx} \{ \psi_{0}(x, E_{n}) {\bar \psi}_{0}(x,E_{n}+t) 
\theta(x)^{-1} \}; \label{v2} \end{eqnarray} \begin{eqnarray} \psi(x,E) = 
\psi_{0}(x,E) - \\ t {\bar \psi}_{0}(x,E_{n}+t) \theta(x)^{-1} \int^{x} 
\psi_{0}(y, E_{n}) \psi_{0}(y,E)dy; \label{psi2} \end{eqnarray} 
\begin{eqnarray}
\psi(x,E_{n}+t) \enskip =
\enskip \psi_{0}(x, E_{n}) \theta (x)^{-1}.
\label{psit}
\end{eqnarray}
It appears that our ability to shift isolated bound state
energy levels (over the $E$-scale) in the infinite potential well on a 
finite interval allows us to move some chosen upper (lower) boundaries of 
the spectral zones (bands) of periodic structures keeping infinite number 
of other upper (lower) boundaries unperturbed.

  Let us consider  instructive particular examples.
The above formulae can be periodically continued and used for 
spectral control of the initial Dirac comb (equidistant  $\delta 
$-potential barriers or wells $\sum_{m=-\infty}^{\infty}
\stackrel{\circ}{V} \delta (x-m\pi)$. To be more precise, we 
get the following periodic potential: 
$\sum_{m=-\infty}^{\infty}\stackrel{\circ}{V} \delta (x-m\pi) + V^{per}(x), 
\enskip V^{per}(x+l\pi)=V_{2}(x), \enskip l=0; \pm 1; \pm 2;...$.
In Fig.1a transformations of  spectral zones in the case 
$\stackrel{\circ}{V}=4$ corresponding to the successive shifts up of second 
level $\stackrel{\circ}E_{2}=4 \rightarrow E=9 $ of the auxiliary problem 
are demonstrated. Pay attention to the merging of the second and third 
allowed zones at special value of $\Delta E=1$ and their separating  after 
further moving $E_{2}$ up.

It is remarkable that at first ($\Delta E \le 1$) the lower boundary of the 
upper neighbor allowed zone goes down to meet the approaching level $E_{2}$ 
until the second and the third allowed zones merge. 

Compare it with the opposite situation on (in) Fig.1b when $E_{2}$ is
shifted down, moving away from the third level. Then the zone 
below the fixed level $E_{3}$ shrinks  and goes up until its complete 
disappearance. The second zone at $\Delta E=-1.8$ merges with the first 
zone and after that the first zone recoils down.

These points of zone junction are  
especially interesting. The phenomenon of zone merging can be clarified
if we consider the continuous transformation of Dirac comb with $\delta
$-barriers into one with $\delta $-wells. In the first case the allowed
zones have the auxiliary levels as upper boundaries and in the second as 
lower ones \cite{Flugge}. For zero $V_{\delta }$ all the gaps between zones 
disappear while transiting from repulsion to attraction. Then the 
auxiliary levels $E_{n}$  are simultaneously upper
boundaries for the upper $n$th allowed zones and lower boundaries for the
lower $n-1$th zones  and there is the continuous spectrum from $E=0$ to
$E \rightarrow \infty$.  This allows 
us to suppose that in the case of non-zero potentials there is  "balance"
between attraction and repulsion only in the vicinity of $E$-point of zone 
junction where the $E_{2}$ level becomes a "boundary" for both neighbor 
zones.

When shifting $E_{2}$ further upward there occurs the transition to the 
regime of a local prevalence of the attraction and  $E_{2}$ ceases 
being the upper boundary of the second zone. This zone separates 
from its previous upper boundary $E_{2}$ and recoils downward.
 
The reappearance of the vanished gap between 2nd and 3rd zones 
can be explained as follows.  During the further shifting the $E_{2}$ level 
up through the balance point, the local predominance  of repulsion is 
changed by the prevalence of attraction.  So the level  $E_{2}$ becomes only 
the lower boundary of the upper zone abandoning the lower zone which moves 
down restoring the just disappeared gap. The same effects can be 
observed in Figs.1a,c.

 Let us shift auxiliary energy levels starting from free motion 
without Dirac combs (continuous spectrum with all gaps collapsed).
Then the experience we have already got allows one to predict the qualitative
zone movement. For this it would be useful consideration of right 
sides of Figs.1a-c  after the first zone merging.

Now we shall consider another example of zone control. Below are 
given the formulas for the potential and wave functions transformations 
corresponding to variations of spectral weight factor $c_{n}$ of a chosen 
bound state at the energy $E_{n}$ (see a general case in 
\cite{Obed,IP,Korob}):  Let $\stackrel{\circ}{\psi }_{m}(x)$ be an 
eigenfunction of the initial Hamiltonian corresponding to $m$th energy 
eigenvalue, $\stackrel{\circ}{c}_{m}$'s are the initial spectral 
weight factors.  Then changing $\stackrel{\circ}{c}_{m} \to c_{m}$ gives 
\begin{eqnarray} \psi_{n}(x)=\stackrel{\circ}{\psi }_{n}(x)+ 
\frac{(1-c_{m} ^{2}/\stackrel{\circ}{c}_{m}\!^{2}) \stackrel{\circ}{\psi 
}_{m}(x)} {1-(1-c^{2}_{m}/\stackrel{\circ}{c}_{m}\!^{2}) \int 
\limits_{0}^{x}\stackrel{\circ}{\psi }_{m}\!^{2}(y)dy} \nonumber \\
\times \int \limits_0^x 
\stackrel{\circ}{\psi }_{m}(y)\stackrel{\circ}{\psi }_{n}(y)dy, 
\label{psi} 
\end{eqnarray}
\begin{eqnarray}
V(x)= \stackrel{\circ}{V}(x) \nonumber \\
+ 2\frac{d}{dx}\frac{(1-c_{m}^{2}/\stackrel{\circ}{c}_{m}\!^{2})
\stackrel{\circ}{\psi }_{m}\!^{2}(x)}
{1-(1-c_{m}^{2}/\stackrel{\circ}{c}_{m}\!^{2})
\int \limits_0^x \stackrel{\circ}{\psi }_{m}\!^{2}(y)dy}.
\label{V}
\end{eqnarray}
The transformed solution $\psi_{n}(x)$ differs from the initial one by 
that varying the parameter $c_{n}$ changes the space distribution of wave 
function (localization) \cite{Sof,PoshTrub,ZS}.

We need not restrict ourselves to the discret spectral points of
the auxiliary infinite rectangular well and zero boundary conditions at 
  its walls. In usual periodical problems there are  no 
impenetrable walls. So, arbitrary homogeneous boundary 
conditions apply for our goals.  Corresponding corrections for the 
above formulas can be easily derived \cite{Obed,IP}

 It turns out  that the algorithms of shifting localization 
of eigenfunctions over the  configurational axis x  result in changing the 
degree of forbiddenness  of the spectral gap at the chosen energy point, 
see  Fig.1d. It can be easier understood in the case of symmetrical initial
potentials with the (anti)symmetrical eigenfunctions. Then variation of 
$c_{n}$  violates the symmetry which results in exponential growth of 
the periodically continued solution swinging. This is
peculiar to the forbidden zone. The more the change in $c_{n}$ the greater
the degree of forbiddenness (i.e., $c_{n}$ is a `lever' of direct control). 
The same is true for the general boundary conditions, i.e., arbitrary 
energy point in the forbidden zone.  In particular, this drastically 
(`continuous' multiplicity) extends the class of exactly solvable models, 
which may be of great value in clarifying different aspects of periodic 
system theory.   

The transformations based on shifting a chosen energy level $E^{I}_{\mu }$ 
in one of two auxiliary spectra corresponding to the same system are 
equivalent to changing infinite spectral weight factors of another 
spectrum, e.g., increasing  $c^{II}_{n}; n> \mu  $ and decreasing 
$c^{II}_{n}; n< \mu $ \cite{Dask}. We have not still analyzed the influence 
of such shifts on band spectral structure. 

Just before  the submission of this paper we revealed unexpected
symmetry. The shifts of "nonphysical" levels $E  \rightarrow E \pm \Delta 
E$ in the infinite  rectangular well give the same deformation (we started 
from free motion) of the band structure of the periodically continued 
potential for arbitrary $\Delta E$ and $E$, which does not  violate the  
rule: no crossing the levels of the physical spectrum while shifting energy 
level.  It is an  intriguing   open problem to explain the phenomenon of 
coincidence of band structures  for different periodic potentials, which  
(up to our knowledge) was never mentioned before.

{\it Conclusions } --
 New status of quantum mechanics is due to achievements of the inverse
problem approach \cite{Obed,IP}.
Instead of about ten exactly solvable models which serve as a basis of
the contemporary research and education there are infinite (!) number, even
{\bf complete} sets of such models. So, the whole quantum mechanics is
embraced by them.  They correspond to all possible variations of spectral
parameters which determine all properties of quantum systems.  There
appears a possibility to change at wish quantum objects by variation of
these parameters as control levers and examine quantum systems in different
thinkable situations. The regularities revealed by computer visualization
of these models were reformulated into unexpectedly simple universal rules
of arbitrary transformations and what is more, their elementary
constituents were discovered.  Of these elementary `bricks' it is possible
in principle to construct objects with any given properties.
The  embedding  of potentially real objects into the continuum of all the
thinkable exactly solvable models allows to unify the whole picture, gives a
notion of the connections between the systems considered before as
completely different ones. This simplifies the science and gives a
significant economy of memory to be used for our further creative activity.

  The well-known finite-gap potentials \cite{March,Levit} might serve as a 
good example of corresponding exact models.  At the same time these 
potentials have the obvious fault that they are not flexible enough:  they 
have `frozen' distribution of forbiddeness index.  

\newpage
FIGURE CAPTIONS

Fig.1a Shifting up  the upper boundary
of the second allowed zone for the Dirac comb (periodical $\delta
$-barriers). This boundary coincides with the second energy level in the
auxiliary well with impenetrable walls at the ends of finite 
period interval. It pools at first its zone while moving upward.  And the 
lower boundary of the neighbor upper 3rd zone is lowering toward it
until at $\Delta E=1$  both neighbor zones merge.  The second energy 
level of the transforming auxiliary well becomes the lower boundary of the 
upper zone at this moment.  After that the zones are separating again 
$\delta E=2,3,3.9$, but the shifted level now remains with the upper zone.  
Its motion up squeezes the upper zone (between two levels) until 
annihilation, because its upper boundary $E=9$ remains  unchanged according 
to the algorithm of shifting the only one energy level.

  Fig.1b Shifting down the upper boundary of the 2nd allowed zone of
the Dirac comb of  $\delta $-barriers.  This boundary pushes downward the 
2nd zone before it until its lower boundary touches the lower first
allowed zone which fixed boundary $E=1$ becomes at this moment also the 
lower boundary of the zone coming from above. After the zone merging the 
lower zone separates from the upper one and goes downward while the 
upper zone is squeezed between its boundaries: the fixed  $E=1$ and 
lowering second energy level of the auxiliary well.

Fig.1c Shifting up the lower boundary of the second allowed zone of
Dirac comb  $\delta$-wells (ground state of the auxiliary eigenvalue
problem on the period).  This boundary pushes the zone upward from below 
until its upper boundary achieves the second fixed energy level 
$E=4$.  At this moment the spectral gap between the two zones 
disappears.  After that the squeezing of the lower zone  begins similarly 
to the case  of Fig.2. Simultaneously, the upper zone separates and goes 
upward restoring the just disappeared lacuna. The degeneration of auxiliary 
energy levels is followed by their effective annihilation \cite{IP,ZS} 
and by the collapse of all the allowed zones.

Fig.1d (upper part)  Changes in zone structure corresponding to 
relative variations of spectral weight factors $c/\stackrel{\circ}{c}$ at 
energy point $E=2$.  The initial periodic potential $\stackrel{\circ}{V} 
\delta (x-n \pi )$ has Dirac comb peaks with the strength
$\stackrel{\circ}{V}=4$ and period $\pi $. The wave function at $E=2$  is on 
each period an eigenfunction of the eigenvalue problem with 
the boundary conditions on the edges of a period specified so that 
the auxiliary discrete energy level just coincides with the chosen point 
$E=2$.  (Lower part) Changes of the imaginary part of quasi-momentum  $Im 
K(E)$ which characterize the degree of forbiddeness, the index  of 
exponential swinging of solutions.
\end{document}